\documentclass[twocolumn]{htl-author}

\usepackage{url}            
\usepackage{booktabs}       
\usepackage{amsfonts}       
\usepackage{nicefrac}       
\usepackage{microtype}      
\usepackage{graphicx}
\usepackage{subfig}
\usepackage{multirow}

\usepackage[letterpaper]{geometry} 
\usepackage[english]{babel}

\usepackage{amsmath}
\usepackage{graphicx}
\usepackage{subfig}
\usepackage[colorinlistoftodos]{todonotes}
\usepackage{enumitem}
\usepackage{float}
\usepackage{multicol}
\usepackage{empheq}
\usepackage{multirow}

\begin{document}

\title{Severity Detection Tool for Patients with Infectious Disease}
\small
\author{Girmaw Abebe Tadesse$^{1*}$,  Tingting Zhu$^{1*}$,  Nhan Le Nguyen Thanh$^{2}$, Nguyen Thanh Hung$^{2}$, Ha Thi Hai~Duong$^{3}$, Truong Huu Khanh$^{2}$, Pham Van~Quang$^{2}$, Duc Duong Tran$^{3}$, Lam Minh Yen$^{4}$, H Rogier Van~Doorn$^{5,6}$, Nguyen Van Hao$^{3}$, John~Prince$^{1}$, Hamza~Javed$^{1}$, Dani Kiyasseh$^{1}$, Le Van Tan$^{4}$, Louise~Thwaites$^{4,6}$ and  David A. Clifton$^{1}$\footnote{* Joint first authors}\footnote{This paper is a preprint of a paper submitted to Healthcare Technology Letters. If accepted, the
copy of record will be available at the IET Digital Library}
} 

\address{$^{1}$Institute of Biomedical Engineering, University of Oxford, Oxford, UK\\
$^{2}$Children's Hospital Number 1, Ho Chi Minh City,  Vietnam\\
$^{3}$Hospital for Tropical Diseases, Ho Chi Minh City, Vietnam\\
$^{4}$Oxford Clinical Research Unit, Ho Chi Minh City, Vietnam\\
$^{5}$Oxford University Clinical Research Unit, Hanoi, Vietnam\\
$^{6}$Centre for Tropical Medicine and Global Health, Oxford University, UK\\
Corresponding Email:\textcolor{blue}{girmaw.abebe@eng.ox.ac.uk}

}


\abstract{\textbf{Hand foot and mouth disease (HFMD) and tetanus are serious infectious diseases in low and middle income countries. Tetanus in particular has a high mortality rate and its treatment is resource-demanding. Furthermore, HFMD often affects a large number of infants and young children. As a result, its treatment consumes enormous healthcare resources, especially when outbreaks occur. Autonomic nervous system dysfunction (ANSD) is the main cause of death for both HFMD and tetanus patients. However, early detection of ANSD is a difficult and challenging problem.  In this paper, we aim to provide a proof-of-principle to detect the ANSD level automatically by applying machine learning techniques to  physiological patient data, such as electrocardiogram (ECG) and photoplethysmogram (PPG) waveforms, which can be collected using low-cost wearable sensors. Efficient features are extracted that encode  variations in the waveforms in the time and frequency domains. A support vector machine is employed to classify the ANSD levels. The proposed approach is validated on multiple datasets of HFMD and tetanus patients in Vietnam. Results show that encouraging performance is achieved in classifying ANSD levels. Moreover, the proposed features are simple, more generalisable and outperformed the standard heart rate variability (HRV) analysis. The proposed approach would facilitate  both the diagnosis and treatment of infectious diseases in low and middle income countries, and thereby improve overall patient care.}}

\maketitle

	\section{\textbf{Introduction}}
	
	Infectious diseases, such as tetanus and hand foot and mouth disease (HFMD),	still pose life-threatening risks to patients in low and middle income countries~[1]. Tetanus is a severe disease, often necessitating lengthy hospital treatment (up to six weeks), which was estimated to have  caused 48-80,000 deaths in 2015~[2]. It tends to affect the poorest in society in low and middle income countries where  unvaccinated individuals, particularly manual workers and farmers, are at high risk of contracting it~[1,~3,~4,~20]. A recent study showed that tetanus prevalence is still high in a part of Asia, and that it is associated with high morbidity and mortality rates~[17,~20,~25].
	
	Comparatively, HFMD is typically a benign self-limited illness in infants and young children.  In recent years, large outbreaks have been reported in the Asia Pacific region, affecting millions of children~[5,~6,~18]. For example, $90\%$ of HFMD incidents in China occur among children under the age of 5 years~[18]. Although most HFMD cases are mild, a small number of affected children  progress rapidly to severe or fatal manifestations of the disease. Moreover, survivors may still be afflicted with neurocognitive impairments later in life, despite having apparently full recovered from severe HFMD~[18]. Inability to predict those who will progress to severe cases means that huge numbers of children are admitted to hospital as a precautionary measure, placing an enormous burden on healthcare systems~[5,~6,~7]. 
	
	Autonomic nervous system dysfunction (ANSD) is the main cause of death in the aforementioned infectious diseases~[2,~6,~7,~22,~23]. It is not clinically apparent in the early stages of disease, but once it is established, treatment is challenging and, in the case of HFMD, deterioration can occur rapidly. In tetanus, early diagnosis may enable preventative intervention and allow differentiation from other causes of tachycardia and hypertension. 
	
\begin{figure}[h]
	\centering
	{\includegraphics[scale=0.24]{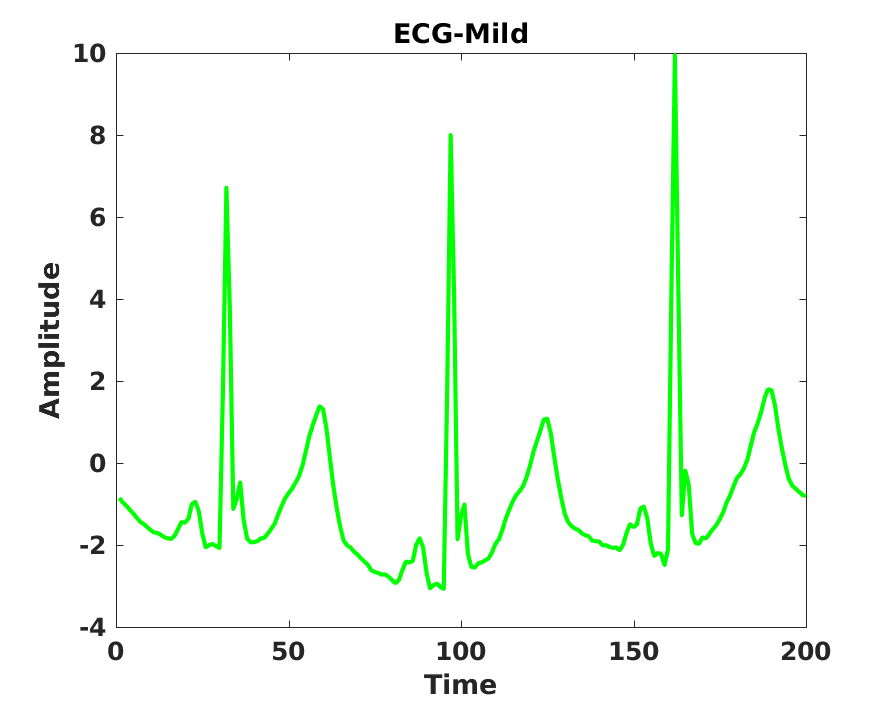}}{\includegraphics[scale=0.24]{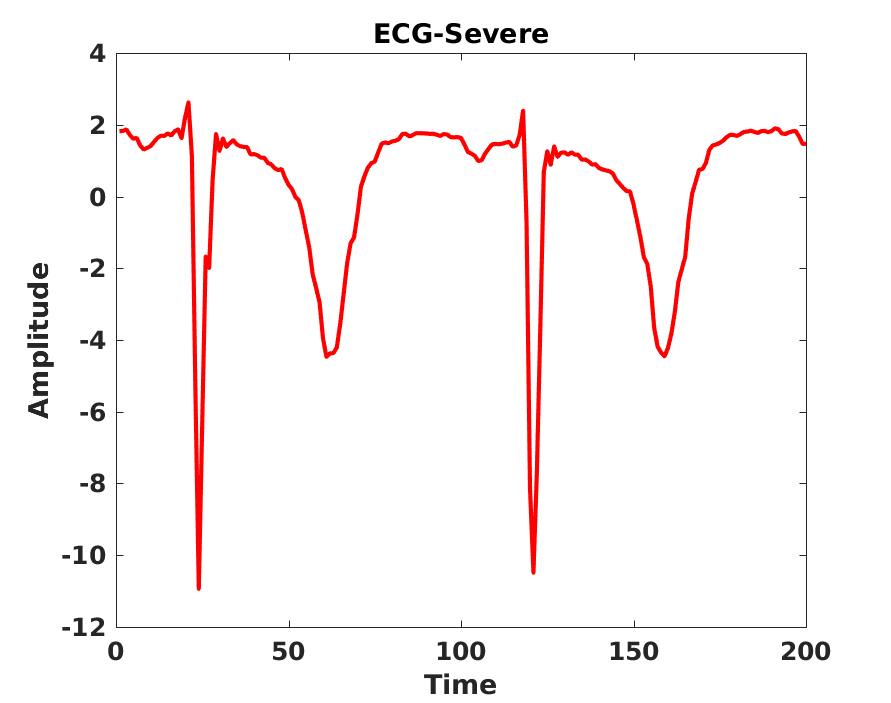}}
		\caption{Example of ECG waveforms (amplitude vs. time) from  two randomly selected tetanus patients}\label{fig:mild_severe}
		
	\end{figure}

    Data-driven approaches have been employed to assist clinicians making informed decisions during the diagnosis of infectious diseases~[8,~9,~15,~16].  The physiological data from patients (see Fig.~\ref{fig:mild_severe})  mainly include electrocardiogram (ECG)~[8,~21] followed by photoplethysmogram (PPG)~[9,~16] waveforms. 

    Existing methods are mainly focused on heart rate variability (HRV) analysis based on a prior detection of morphological features~[8,~9,~15,~16,~21]. This means ECG-based features were derived from P-wave, R-peak, T-wave and the PQ, QRS and
ST segments. Similarly, PPG-based features were derived from the systolic and diastolic
segments.
However, morphology-based feature extraction requires, in addition to more computational resources, domain-specific knowledge; and these features are hardly transferable across different physiological waveforms (e.g. ECG, PPG, and IP). Furthermore, these morphological features could easily be affected by noise and motion artefacts, especially when wearable devices are employed and/or the patients are children who are prone to make random movements. The traditional approach that employs a specific clinical monitor or Holter device has been found to have limitations in clinical practice, especially with small children in the out-patient setting~[7,~24]. As a result, these features are less robust and have limited generalisability across  variations in patient characteristics and device specifications.

\begin{figure*}[t]
		\includegraphics[scale=0.25]{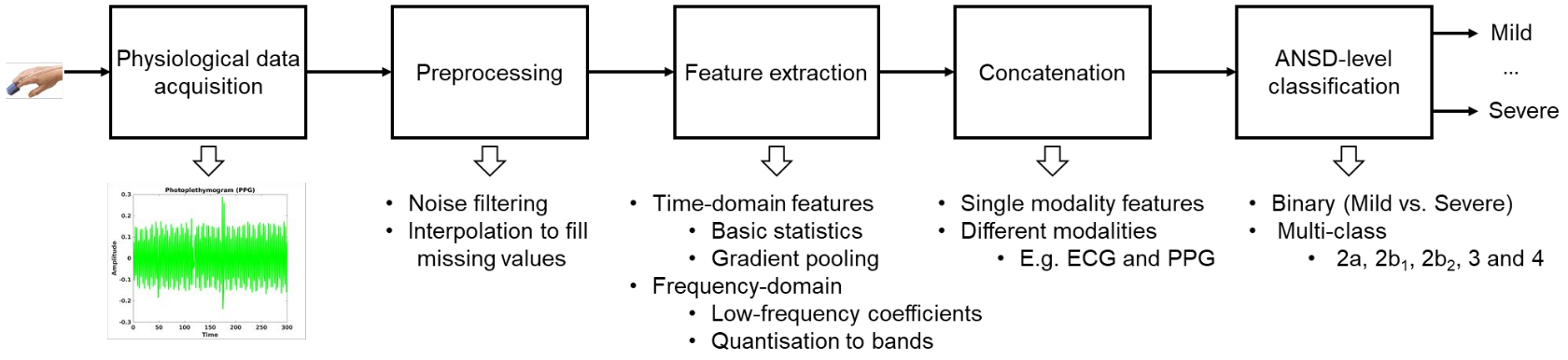}
		\caption{Block diagram of the proposed approach}\label{fig:bdg}
		\end{figure*}

	In this paper, we present our preliminary work on automatic ANSD detection using multivariate physiological data  collected from tetanus and HFMD patients in Southern Vietnam (see Fig.~\ref{fig:bdg}). The proposed approach could be integrated into the clinical pathway to provide a low-cost care tool to triage patients.  We collected physiological waveforms from children using wearable devices, which are low-cost, non-invasive and easy to wear. In addition, these devices are cost-effective for resource-limited settings such as  low and middle income countries~[9,~10]. After data collection, feature extraction is applied to encode the variability of these waveforms both in the time and frequency domains. The proposed features are designed to be simple and generalisable across different physiological waveforms (e.g. PPG and ECG) without a prior detection of domain-specific morphological variations. Later, a state-of-the-art classifier is applied to discriminate the ANSD levels of patients. We also applied feature-level fusion when multivariate data  was available. This automatic tool for ANSD detection aims to support efficient allocation of resources, and hence improve patient care. In addition, as patients with these diseases are often given antibiotics, the creation of a robust and reliable detection tool may also reduce unnecessary use of antibiotics and therefore limit antimicrobial resistance.

	\section{\textbf{Related Works}}
	As the heart is under autonomic nervous system control, changes in beat-to-beat variability of the heart rate, detected by the ECG, have been linked to changes in autonomic system balance [8].
	
	Lin et al.~[15] showed that patients with different stages of HFMD experienced different levels of central nervous system complications, which was reflected by their HRV measures. Though HRV has been principally inferred from ECG signals, PPGs could be promising alternatives as existing methods in the literature reported HRV parameters derived from PPG had high correlation with those derived from ECG~[9,~16]. It is encouraging to be able to carry out PPG-based HRV analysis, as ECG acquisition is relatively complex, requiring electrodes to be mounted on specific anatomical positions, which may cause skin irritations and be less practical in non-clinical settings~[9,~16].

		However, existing HRV-based approaches to evaluate autonomous dysfunction mainly require the detection of morphological shapes and features (e.g. QRS complex and  RR intervals)~[8,~9,~25], which incurs an additional pre-processing step.  In addition, these features are not generalisable across different vital signs which follow different morphological shapes that could be easily affected by artefacts. Furthermore, HRV parameters obtained with non-linear modes, such as standard deviations of short and long diagonal axes in the Poincare plot (SD1 and SD2), necessitate additional computational cost.

The significance of our approach lies in the development of more generic features rather than domain-specific ones (e.g. PQRST characteristics for EEG and systolic and
diastolic features for PPG). That means there is no need
for prior detection of these morphologies. As a result, our approach can be employed across a variety of
modalities that are time-series bio-signals. Bio-signals are easier and cheaper to collect, typically involving less obtrusive collection compared to clinical tests. This approach could therefore also enable remote monitoring of patients by
their care giver, using existing wearable sensor technologies.
	
	\section{\textbf{Proposed Method}}\label{sec:method}

	The proposed approach consists of pre-processing, feature extraction and classification stages as shown in Figure~\ref{fig:bdg}. Physiological data collected using wearable sensors are often susceptible to noise and movement artefacts. Hence, a high pass filter followed by a Gaussian filter is applied to mitigate these challenges during the pre-processing stage. Moreover, we aim to ensemble multiple but simple  time- and frequency-domain features to form a more robust feature set overall. In addition, the gradient-based feature extraction (described later) helps to further encode noise-free features. Feature extraction is  applied to each window of data points segmented from a waveform. The window duration determines the number of samples extracted from a continuous waveform.
	
	Given a time-series of physiological data, $\textbf{x}=(\textbf{x}_n)_{n=1}^L$, where $L$ represents the number of data points in a window, we propose to extract both time- and frequency-domain features. The time-domain features are further grouped into gradient- and non-gradient-based features.  Non-gradient-based time-domain features encode the basic statistics  of the signal, such as \textit{minimum, maximum, median, mean, standard deviation, energy, kurtosis} and \textit{zero-crossing}~[11]. \textit{Energy} is obtained as $\sum_{n=1}^L\textbf{x}_n^2$. \textit{Kurtosis}, $k_x$, measures the deviation of a signal distribution from a Gaussian distribution, that is $k_x=L\frac{\sum_{n=1}^L(\textbf{x}_n-\mu)^4}{(\sum_{n=1}^L(\textbf{x}_n-\mu)^2)^2},$ where $\mu$ is the mean of $\textbf{x}_n$.  \textit{Zero-crossing} refers to the number of times a signal amplitude crosses the zero-magnitude threshold and encodes oscillation characteristics.
	
	Gradient-based features help to extract more dynamic information in the time-domain~[12]. The gradient is computed by applying first-order derivative, i.e.~$\textbf{x}'_n~=~\textbf{x}_{n+1}-\textbf{x}_n$. Two specific gradient pooling features, count ($\textbf{h}_x$) and sum ($\textbf{s}_x$) of the gradient histogram are extracted. Count pooling counts positive ($\textbf{h}^+_x$) and negative gradients ($\textbf{h}^-_x$), whereas sum pooling sums all positive ($\textbf{s}^+_x$) and negative gradients ($\textbf{s}^-_x$). For example, $\textbf{h}^+_x$ of $\textbf{x}_n$ is computed as $\textbf{h}^+_x= \sum_{n=1}^{L-1} s(\textbf{x}'_n),$ where 	\[
	s(\textbf{x}'_n)= 
	\begin{cases}
	1,& \text{if } \textbf{x}'_n\geq 0\\
	0,              & \text{otherwise}. 
	\end{cases}
	\]

	Frequency-domain features provide more detailed dynamic information using the fast Fourier transform (FFT). 

 The frequency-domain features can be grouped into two groups: low-frequency ($\textbf{f} (_{l}$) and whole-frequency features ($\textbf{f}_{w}$). Low-frequency features contain the magnitude of $N_{l}$ low frequency coefficients after the Fourier transform. Full-frequency group includes the sum of frequency response magnitude of frequency bins clustered into $N_{b}$ consecutive bins. The significance of the frequency features is as follows. $\textbf{f}_{l}$ contains high-resolution low-frequency characteristics, as much of the energy rests in this frequency band. On the other hand, $\textbf{f}_{w}$ contains the whole spectrum (both the low and high frequency patterns) with lower resolution. This is motivated by the need to include the high frequency characteristics and their comparison with lower frequency ones.  As a result, $\textbf{f}_{w}$ encodes the complete frequency spectrum compared to $\textbf{f}_{l}$. We tend to cluster the frequency components into  bins to have lower resolution since higher resolution might result in unnecessarily long feature dimensions.
	
 Let $\textbf{f}_x=\mathcal{F}(x_n)$ be the frequency response of $x_n$, $\textbf{f}_l=\{\textbf{f}_x(c)\}_{c=1}^{N_l}$ and the $\textbf{f}_{w}$ feature (with $N_b$ bands) is computed as
	\small$\textbf{f}_{w}(j) = \sum_{l=\sigma_i}^{\sigma_f}\textbf{f}_x(l),$
	where \small$$j\in [1,N_b],
	\sigma_i=1+\frac{(j-1)*L}{2N_b},
\sigma_f=1+\frac{j*L}{2N_b}.$$

The final feature vector is obtained using a simple concatenation, $\mathcal{C}(\cdot)$, of both time- and frequency-domain features into a single vector. Given two feature vectors, $\textbf{f}_1\in \mathbb{R}^{d_1}$ and $\textbf{f}_2\in \mathbb{R}^{d_2}$, their concatenation $\textbf{f}_c=\mathcal{C}(\textbf{f}_1, \textbf{f}_2)$ results in $\textbf{f}_c \in \mathbb{R}^{d_c}$, where $d_c=d_1+d_2$.  Similar concatenation approach is applied for features from different modalities, (e.g. ECG and PPG). Finally, we employ support vector machines (SVM) to classify the ANSD severity levels.

\section{Complexity Analysis}
In this section, we present the complexity analysis of the feature extraction step, per feature type, in the proposed framework.
Given a time-series signal of $L$ time steps, the computational complexity of the majority of the time-domain features (e.g. mean and median) have linearly growing complexity, i.e. $\mathcal{O}(L)$. However, the gradient features may have additional complexity of $\mathcal{O}(2L)$ due to the first-order derivative and the summing/counting of positive and negative gradients.
The Fourier transform for the frequency-domain features ($\textbf{f}_l$ and $\textbf{f}_w$) pose a computational cost of $\mathcal{O}(L\log(L))$ associated with the FFT computation.  
In addition, we provide Table~\ref{table:computationTime} that summarises the wall-clock computation time elapsed for the extraction of the proposed features for a randomly selected PPG signal that is $\approx 5$ minutes long. The whole feature extraction takes $\approx 21.15$ ms, of which time-domain features elapse $\approx 17$ ms and frequency-domain features elapse $\approx 4.15$ ms. The experiments were conducted using Matlab2017a, Intel(R) Xeon(R) CPU E5-1630 v3 @ 3.70GHz, Ubuntu 16.04 OS and 32GB RAM.

\begin{table}
	\centering
	\caption{Summary of wall-clock time elapsed for the\\ computation of time and frequency features, experimented\\ on a randomly selected $\approx 5$-mins PPG signal.} \label{table:computationTime}
	\resizebox{0.6\linewidth}{!}{%
		 \begin{tabular}{llc}
			\hline
		Feature group & Feature & Elapsed time (ms) \\ \hline \hline
			\multirow{9}{*}{Time}& Mean & 3.1   \\
			& STD & 1.6  \\
			& Zero-crossing & 1.2  \\
			& Minimum & 0.1  \\
			& Maximum & 0.1\\
			&Median & 3.0 \\
			& Energy & 0.3 \\ 
			& Kurtosis & 3.2   \\
			& Gradient & 4.4 \\ \hline
			\multirow{2}{*}{Frequency}& Low Freq. &2.9 \\
			& Whole Freq. & 4.1 \\
			\hline
		\end{tabular}
	}
\end{table}
			
					



	\section{\textbf{Data Collection}}

	We validate the proposed approach on datasets of HFMD and tetanus patients admitted in hospitals in  Vietnam\footnote{The study was approved by the relevant Ethical Committees and carried out in line with the declaration of Helsinki.}.  The HFMD dataset was collected from Children Hospital No. 1, Ho Chi Minh City, and contains 74 HFMD patients, with a majority of children less than three years old. Commercial devices such as E-patch\footnote{\url{epatch.madebydelta.com}} were used to collect ECG (256 Hz) waveforms in the HFMD dataset. Specifically, 24 hour-patch ECGs are recorded at least twice, when patients are admitted to the infectious disease department and on the penultimate day of hospitalisation. We used the clinical diagnosis of the HFMD patients (based on the clinical grading system developed by the  Vietnamese Ministry of Health) as the ground truth and it contains five levels (in the increasing order of severity): $2a(33)$, $2b_1(9)$, $2b_2(11)$, $3(20)$ and $4(1)$. The number of patients per class is shown in brackets.	 There is a significant imbalance in the number of cases (patients) of ANSD severity levels. Therefore, we merged $2b_1$ and $2b_2$ into a single class. Similarly, level-$3$ and -$4$ were also merged together.
	
	The tetanus dataset contains ECG, PPG and IP waveforms, each lasting up to 24 hours, collected from a total of 10 patients (four moderate disease, Ablett Grade 3 and six severe disease Ablett Grade 4) admitted to the intensive care units (ICU) in the Hospital for Tropical Diseases, Ho Chi Minh City. The sampling rates of ECG, PPG and IP waveforms are 300~Hz, 100~Hz and 25~Hz, respectively. ECG and PPG were time synchronised and recorded from all the patients, which makes the feature-fusion of these different modalities easier. However, it is worth noting that IP signals are missing in four subjects. 
	A  Datex  Ohmeda  monitor  and  a  pulse  oximeter were  employed  for  data  acquisition. VS Capture software [14] was used to download the signal from the monitor. 
	The clinical diagnosis of tetanus patients (i.e. moderate or severe) is used as a ground truth for the experiments. {For a window duration of 5 minutes, the number of samples extracted from each modality are $3,077$ (ECG), $3,070$ (PPG) and $1,895$ (IP). From HFMD dataset, a total of $60,373$ samples are extracted from the ECG signal.}


	\section{\textbf{Parameter Setup}}
	During the feature extraction step, we set the window duration to be at least five minutes, similar to the duration in the clinical baseline method~[8] extracted using a publicly available software solution~[13].  {The baseline method was selected because it has been a gold standard for many existing works that focused on HRV analysis~[9,~15,~16,~20]. Recently, a similar method has been used to study HRV among tetanus patients.} {We set the high-pass and low-pass cutoff frequencies to $0.05 $ Hz and $150$ Hz, respectively, in the pre-processing step to filter out artefacts in the physiological signals.}   A temporal resolution of two is applied to extract gradient pooling features. A temporal resolution refers to the number of chunks the original sample is divided into. E.g. given a 5-minute long waveform, a temporal resolution of two means divide the signal into two chunks (each 2.5 minutes long) and extract gradient-based features on each of them. We set $N_l=200$  and $N_b=200$ in order to achieve a balance between higher frequency resolution and smaller feature dimension, i.e. lower values of  $N_l$ and  $N_b$ result in lower frequency resolution but smaller feature dimension, whereas their higher values result in better frequency resolution but longer feature dimension. 
	Both linear and Gaussian kernels are experimented with the SVM-based classification. We split the data to train and test sets with a ratio of $80\%$ and $20\%$, respectively. The classification is repeated 100 times, each with with different initialisation of the classifier, and their average performance is reported (along with the standard deviation across the iterations).
	
	We employ the following performance metrics:  accuracy~(${A}$), precision~(${P}$), sensitivity or recall~(${R}$), specificity~(${S}$) and F-score~(${F_1}$),  defined for a binary classification as follows. 
  $\small
 P = \frac{TP}{TP + FP}, \hspace{0.5cm} R =\frac{TP}{TP + FN}, \hspace{0.5cm} S =\frac{TN}{TN + FP},  
A=\frac{TP + TN }{TP + TN + FP + FN}, \hspace{0.5cm} F_1 = \frac{2 \times P \times R}{P + R} \nonumber
 $where \textit{TP}:~true positive, \textit{TN}:~true negative, \textit{FP}:~false positive, and \textit{FN}:~false negative samples. For example, in Mild vs. Severe classification of tetanus patients, \textit{TP} refers to the number of samples correctly identified as Severe and similar to the
ground truth label; \textit{TN} refers to the number of samples correctly identified as Mild and similar to the ground
truth label; \textit{FP} refers to the number of samples incorrectly classified as Severe but 
labelled as Mild in the ground truth; and FN refers to the number of samples misclassified as Mild but  labelled as Severe in the ground truth. For the HFMD dataset, which involves multi-class classification, an SVM with one-vs-all (OVA) strategy is used. For example, during the
classification of class~2a, samples from this class are positive samples and all the samples
from the remaining classes (i.e. 2b, 3 and 4) are treated as negative samples. For example, during the
classification of class~2a, samples from this class are positive samples and all the samples
from the remaining classes (i.e. 2b, 3 and 4) are treated as negative samples. The performance metrics are initially computed for each OVA classification and the average performance across the classes is reported as a final result.	

		\begin{figure*}
			\centering
			\includegraphics[width=0.3\textwidth]{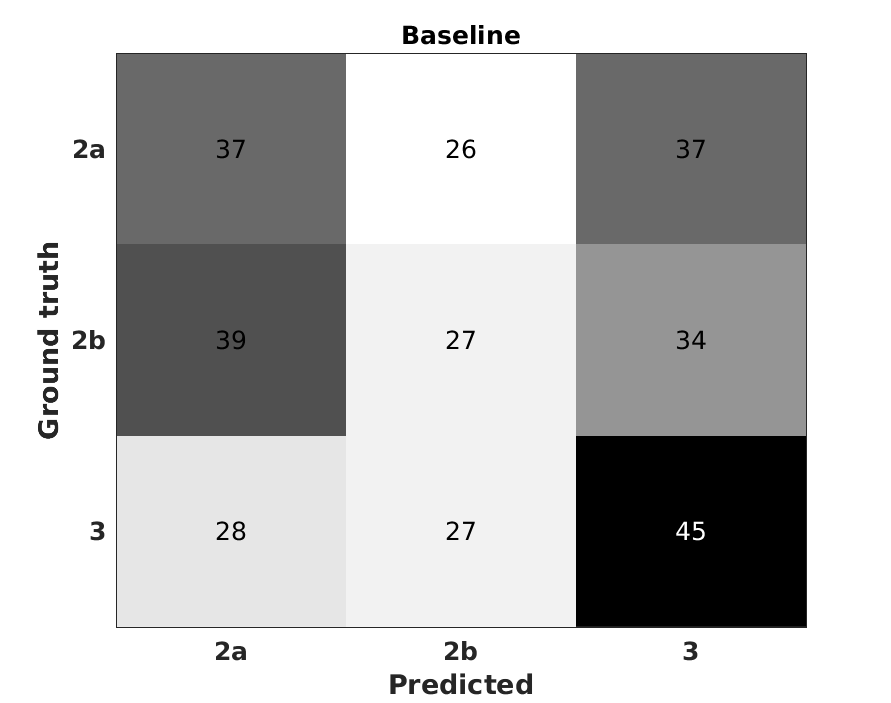}
			\includegraphics[width=0.3\textwidth]{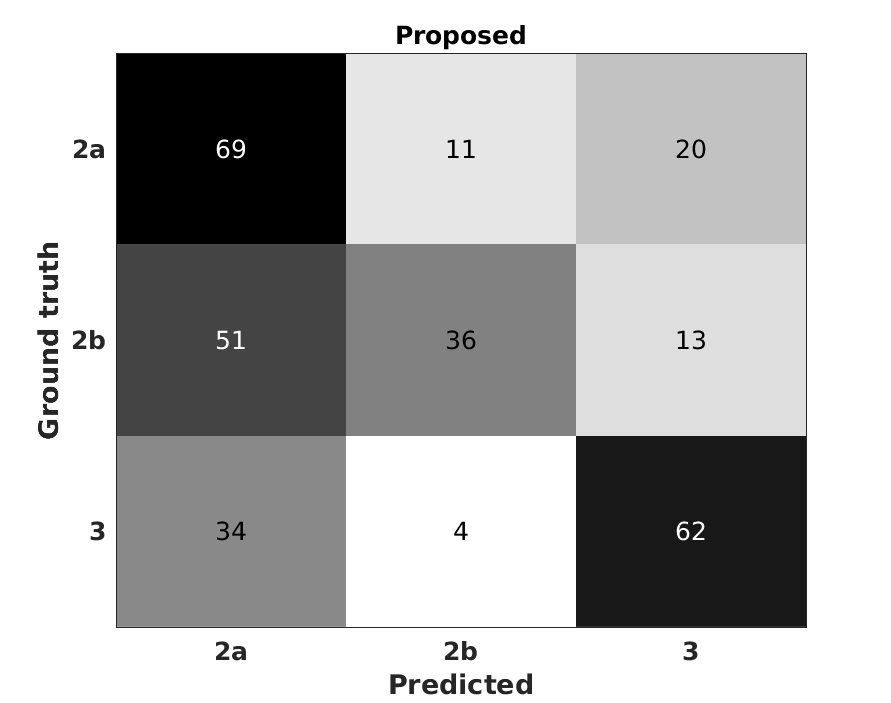}
			\includegraphics[width=0.3\textwidth]{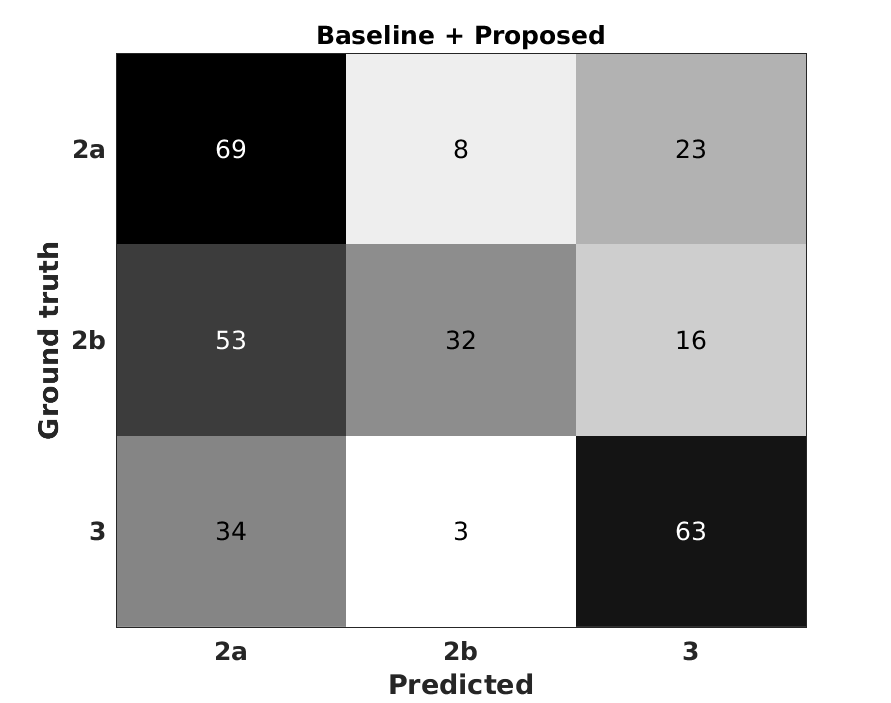}
			\captionof{figure}{Normalised confusion matrices (\%) of baseline, proposed features and their concatenation for ANSD level classification of HFMD patients, with dark colours representing higher magnitudes.}\label{fig:hfmd_results}
		\end{figure*}

\begin{table}[h]
			\centering
			\captionof{table}{ANSD level classification of HFMD patients.}\label{table:HFMD_results}
			\resizebox{1.0\linewidth}{!}{
			\begin{tabular}{lccccc}\\
				&	\multicolumn{5}{c}{SVM ($\%$) - Linear Kernel}  \\ \hline
				Features & A & P & R & S & $F_1$	  \\ \hline
				Baseline [8]& $57.1\pm 0.2$ &  $35.0\pm 0.2$&   $35.2\pm0.2$  & $67.6\pm0.1$  & $34.6\pm0.2$ \\
				Proposed & $64.7\pm 0$ &   $49.1\pm 0.1$ &   $46.9\pm 0.1$ &   $73.4\pm 0$ &   $43.2\pm 0.1$ \\
				Concatenated &  $\textbf{66.9} \pm 0.1$  
				& $\textbf{52.0}\pm0.1$ &  $\textbf{50.1}\pm0.2$ &  $\textbf{75.0}\pm0.1$ &   $\textbf{48.0}\pm0.2$ \\ \hline

				   				&	\multicolumn{5}{c}{SVM ($\%$) - Gaussian Kernel}  \\ \hline
				Features & A & P & R & S & $F_1$	  \\ \hline
                Baseline [8]& $57.7\pm 0.7$ &  $36.2\pm 0.4$&   $36.3\pm0.7$  & $68.2\pm0.4$  & $35.7\pm0.6$ \\
				Proposed & $\textbf{70.9}\pm 0.1$ &   $\textbf{60.6}\pm 0.1$ &   $\textbf{55.9}\pm 0.2$ &   $\textbf{78.0}\pm 0.1$ &   $\textbf{55.7}\pm 0.2$ \\
				Concatenated &  $70.2 \pm 0.1$  
				& $60.0\pm0.1$ &  $54.5\pm0.1$ &  $77.3\pm0.1$ &   $53.9\pm0.2$ \\ \hline
			\end{tabular}
			}
			
		\end{table}

\begin{table}
			\centering
			\captionof{table}{Severity-level classification of tetanus patients.}\label{table:tetanus_results}
			\resizebox{1.0\linewidth}{!}{
			\begin{tabular}{lccccc}\\

   				&	\multicolumn{5}{c}{SVM ($\%$) - Gaussian Kernel}  \\ \hline
				Data & A & P & R & S & $F_1$	  \\ \hline
				Baseline-ECG[8]&$73.9 \pm 0.9$ & $75.48 \pm 1.5$ & $77.5  \pm 0.4$& $67.73  \pm 3.1$ & $76.48 \pm 0.6$ \\ \hline
				IP& $65.7\pm 1.3$ &  $63.2\pm 1.0$&   $94.7\pm0.2$  & $27.8\pm2.8$  & $75.8\pm0.7$ \\
				PPG & $70.2\pm 1.0$ &   $70.4\pm 0.8$ &   $92.6\pm 0.3$ &   $29.5\pm 2.9$ &   $80.0\pm 0.5$ \\
				ECG&  $\textbf{80.2} \pm 0.7$  
				& $\textbf{78.4}\pm0.9$ &  $95.3\pm0.5$ &  $\textbf{53.4}\pm2.5$ &   $\textbf{86.0}\pm0.4$ \\ 
				ECG+PPG&  $78.2 \pm 1.0$  
				& $75.3\pm1.0$ &  $\textbf{98.1}\pm0.0.3$ &  $43.1\pm3.3$ &   $85.2\pm0.6$ \\\hline
			\end{tabular}
			}
\end{table}

	\section{\textbf{Results and Discussion}}
	The proposed approach provides encouraging results in both HFMD (see Table~\ref{table:HFMD_results}) and tetanus (see Table~\ref{table:tetanus_results}) datasets. It is evident from Table~\ref{table:HFMD_results} and Figure~\ref{fig:hfmd_results} that the baseline features, which require detection of QRS complex prior to the feature extraction, fail to discriminate the severity levels of ANSD in HFMD patients. Moreover, the confusion matrices in Figure~\ref{fig:hfmd_results} show a  misclassification of $2a$ and $2b$ classes as there is no well-defined clinical threshold to separate them. The higher classification of class-$3$ to $2a$ than to $2b$ is partly due to the class imbalance, and requires further investigation. We experimented with both linear and Gaussian kernels for the SVM, and Table~\ref{table:HFMD_results} shows that Gaussian kernel performs significantly better than the linear kernel, particularly for the proposed method where about $5\%$ $F_1$-score improvement is achieved using the Gaussian kernel. It is clear that the baseline set of features are less effective at discriminating ANSD levels, and even their concatenation with the proposed features does not provide a significant improvement. The accuracy~(A) and specificity~(S) classification metrics are expectedly higher than the remaining performance metrics, precision~(P), recall~(R) and their $F_1$ score. This is due to the one-vs-all classification strategy employed in the SVM implementation for multi-class classification in the HFMD dataset.
	
	Similarly,  the severity-level classification results of tetanus patients are shown in Table~\ref{table:tetanus_results}. IP achieves the lowest performance compared to ECG and PPG due to the following reasons. First, the number of IP samples is the lowest among all modalities since only six (among ten) subjects had IP waveforms. In addition, the IP waveforms have low sampling rate (25~Hz) in the dataset compared to those of PPG (100~Hz) and ECG (300~Hz). As a result, the IP-based features suffer from the low-temporal resolution of IP waveforms. Higher sampling rate of ECG could also partly explain why ECG performance is better than PPG. In addition, ECG waveforms are relatively stable compared to PPG waveforms as the former are often collected from patient's chest while the latter are collected from motion-prone fingers/toes.  The fusion of features from ECG and PPG waveforms improved the recall  to  $98.1\%$ from their separate recall values of  $92.6\%$ (PPG) and $95.3\%$ (ECG). 
	
     {In clinical practices, 5-min window duration is often applied for HRV. Accordingly, we have used the same duration for comparison with the baseline method in previous experiments. However, we have also experimented the proposed feature extraction method for different window duration (see Table~\ref{table:tetanus_results_1m}.) The results demonstrated that the proposed approach is able to encode time and frequency domain features even for shorter window duration. This is partly due to the repetitive nature of physiological signal characteristics, e.g. QRS complex in ECG. Furthermore, shorter window duration provides higher number of samples for training and hence improved classification performance, as shown in Table~\ref{table:tetanus_results_1m}.}
	
	
	Comparatively, we found that it was difficult to classify the severity levels of HFMD patients, which we hypothesize could be for the following reasons. First, the HFMD dataset was collected from children which are highly likely to move compared to the more static adult tetanus patients. The motion artefacts affect the data quality and degrade the classification performance.  As a result, the features extracted from the PPG waveforms in HFMD patients are less discriminative compared to ECGs shown in Figure~\ref{fig:hfmd_results}.  In addition, a  multi-class classification in HFMD dataset (i.e. three classes) is more challenging than the binary classification problem in the tetanus dataset. 
\begin{table}
			
			\caption{Comparison of 5-minute and 1-minute window\\ duration on the classification of tetanus severity levels.
			}\label{table:tetanus_results_1m}
			\centering
			\resizebox{1.0\linewidth}{!}{
			\begin{tabular}{lccccc}\\
				&	\multicolumn{5}{c}{SVM ($\%$) - Gaussian Kernel}  \\ \hline
				Data & A & P & R & S & $F_1$	  \\ \hline
				IP-5& $65.7\pm 1.3$ &  $63.2\pm 1.0$&   $94.7\pm0.2$  & $27.8\pm2.8$  & $75.8\pm0.7$ \\
				IP-1& $81.2\pm 0.9$ &  $78.0\pm 1.2$&   $93.2\pm0.5$  & $65.5\pm2.7$  & $84.9\pm0.5$ \\ \hline
				PPG-5 & $70.2\pm 1.0$ &   $70.4\pm 0.8$ &   $92.6\pm 0.3$ &   $29.5\pm 2.9$ &   $80.0\pm 0.5$ \\
				PPG-1 & $78.0\pm 0.8$ &   $77.9\pm 0.9$ &   $92.1\pm 0.4$ &   $52.5\pm 2.7$ &   $84.4\pm 0.4$\\ \hline
				ECG-5&  $80.2 \pm 0.7$  
				& $78.4\pm0.9$ &  $95.3\pm0.5$ &  $53.4\pm2.5$ &   $86.0\pm0.4$ \\
				ECG-1&   $\textbf{91.2}\pm0.2$ &  $\textbf{90.7}\pm0.5$ &  $\textbf{96.1}\pm0.3$ &   $\textbf{82.5}\pm1.1$ &$\textbf{93.3}\pm 0.1$ \\\hline
            \end{tabular}
            }

		\end{table}	
	
	
	\section{\textbf{Conclusions}}\label{sec:conclusion}
	We presented our proof-of-principle study to triage patients with infectious diseases (tetanus and HFMD) using low-cost and unobtrusive wearable sensors that collect artefact-prone physiological patient data. For this task, we proposed simple and more generic (across modalities) features to encode the waveform dynamics in time and frequency domains. Our approach was validated on two independent datasets collected from tetanus and HFMD patients in Southern Vietnam.  In addition, the proposed approach provides efficient hospital resource utilisation in low resource-settings, which could in turn help improve overall patient care. The proposed approach still depends on a manual encoding of features. Thus, future works include collecting more patient data and employing data-intensive models, such as deep learning, that generalise better than handcrafted features across variations in patients and acquisition devices.

\ack{{We would like to thank Dr Nguyen Van Vinh Chau, Director of the Hospital for Tropical Diseases, staff in ICU at the Hospital for Tropical Diseases and Children's Hospital Number 1, Ho Chi Minh City.}}

\fundingandinterests{The study was funded by the Wellcome Trust Grants 107367/Z/15/Z and 089276/B/09/7, the Royal Academy of Engineering grant  FoRDF1718$\_$3$\_$19, and the RAEng Frontiers of Engineering for Development under the Global Challenges Research Fund.}








\end{document}